\documentclass[journal]{IEEEtran}
\renewcommand{\baselinestretch}{0.94}\normalsize

 \usepackage{graphics}
 \usepackage{amsmath}
 \graphicspath{{figs/}}
 \usepackage{subfig}
 \usepackage{stfloats}
 \usepackage{amsmath}
 \usepackage{todonotes}

\usepackage{multirow}
\usepackage{hhline}
\usepackage{algorithmic}
\usepackage{array}
\usepackage{subfig}

\begin{document}
\title{Application of ICA on Self-Interference Cancellation of In-band Full Duplex Systems}

\author{Mohammed~E.~Fouda,~
 Sergey~Shaboyan,~
Ayman~Elezabi,~
and~Ahmed~Eltawil,~\IEEEmembership{Senior~Member,~IEEE}
\thanks{Mohammed Fouda, Sergey Shaboyan and Ahmed Eltawil are with Electrical Engineering and Computer Science Dept., University of California--Irvine, CA, USA e-mail: foudam@uci.edu.}
 \thanks{Ayman Elezabi is with the Electronics and Communications Engineering Dept., American University in Cairo, Cairo, Egypt.}
\thanks{The authors gratefully acknowledge support from the National Science Foundation under award number 1710746 as well as the American University in Cairo under Research Support Grant (RSG1-18).}
\thanks{This  work  has  been  submitted  to  the  IEEE  for  possible  publication.Copyright  may  be  transferred  without  notice,  after  which  this  version  mayno longer be accessible}
}


\maketitle

\begin{abstract}
In this letter, we propose a modified version of Fast Independent Component Analysis (FICA) algorithm to solve the self-interference cancellation (SIC) problem in In-band Full Duplex (IBFD) communication systems. The complex mixing problem is mathematically formulated to suit the real-valued blind source separation (BSS) algorithms. In addition, we propose a method to estimate the ambiguity factors associated with ICA lumped together with the channels  and residual separation error. Experiments were performed on an FD platform where FICA-based BSS was applied for SIC in the frequency domain. Experimental results show superior performance compared to least squares SIC by  up to 6 dB gain in the SNR.
\end{abstract}


\IEEEpeerreviewmaketitle
\section{Introduction}
\IEEEPARstart{I}{n-band} full duplex (IBFD) communication systems have witnessed significant progress during the last decade due to developing practical methods to overcome the mixing of the self-interference signal with the signal of interest \cite{liu2015}. Self-interference cancellation (SIC) is generally divided into different stages as follows; 1) Propagation cancellation which is essential to avoid Low Noise Amplifier (LNA) saturation from the interferer. This can be performed via antenna separation, directional antennas, circulators, ... etc. 2) Analog cancellation where the tuned delayed paths are used to subtract the self-interference signal from the received signal, thus reducing LNA saturation and preserving ADC dynamic range, and 3) digital cancellation (DC) to remove any residual interferer signal \cite{liu2015}.

Blind Source Separation (BSS) techniques are widely used to solve cocktail party-type problems to separate linearly and nonlinearly mixed signals \cite{comon2010handbook}. In the case of linear mixtures, the BSS problem is mathematically formulated as follows: 
\begin{equation}
    \boldmath{x=As},
\end{equation}
\noindent where ${\mathbf{A,s}}$ and $\mathbf{x}$ are the mixing matrix, sources, and mixed signals, respectively. BSS methods, therefore, lend themselves naturally to solving the problem of SIC. Independent Component Analysis (ICA) is a powerful tool to separate mixed sources based on their statistical independence and non-gaussianity \cite{hyvarinen2000independent}.  However, BSS techniques suffer two main issues to recover the mixed signals \cite{hyvarinen2000independent}; a) it recovers the mixing matrix with a random permutation, and b) the recovered signals have arbitrary scaling values (scaling ambiguity).           

During the last decade, many ICA algorithms were proposed based on different optimization criteria such as minimizing the mutual information or information maximization or maximizing the non-gaussianity \cite{comon2010handbook}. One of the widely used techniques is Fast Independent Component Analysis (FICA) which works on maximizing the signal non-gaussianity, proposed by Hyvrinen and Oja for real-valued signals \cite{hyvarinen1999fast}. The power of FICA is that it is based on stochastic gradient descent with cubic convergence.

Recently, FICA-based BSS was applied to SIC in full duplex systems  \cite{yang2018digital}. The authors applied their proposed technique in the time domain, albeit without addressing the permutation and scaling ambiguity problems. Furthermore, ideal conditions were assumed where the channel and transceivers nonidealities are ignored such as nonlinearity, phase noise, .. etc. 

Specifically, this letter presents the following contributions:
 \begin{itemize}
     \item A BSS-based method for SIC in In-Band Full Duplex (IBFD) systems applied in the frequency domain is presented for OFDM signals.
     \item A method to estimate the ambiguity factors associated with ICA lumped together with residual separation errors and channels  is presented, avoiding direct channel estimation.
     \item We reformulate the complex mixing problem in order to apply the real-valued FICA scheme.  
      \item  Experimental confirmation is presented by testing the proposed FICA-based BSS on a real IBFD platform showing performance improvements of up to 6 dB in output SINR compared to least-squares SIC.
 \end{itemize}
 
The remainder of this letter is organized as follows: Section II presents the IBFD system model and Least-squares SIC. Section III discusses the BSS formulation of the IBFD problem and the problems associated with the application of BSS in addition to the modified FICA algorithm for resolving the complex mixing problem. Section IV presents the experimental results on the IBFD system. Finally, the conclusion is given in section V. 

\par \textit{Notation:} 
We use $E[.]$ to denote expectation. Frequency domain variables use uppercase letters. Furthermore, bold letters denote vectors.

\section{In-band Full Duplex System}
\begin{figure}[!t]
\centering
\includegraphics[width=0.85\linewidth]{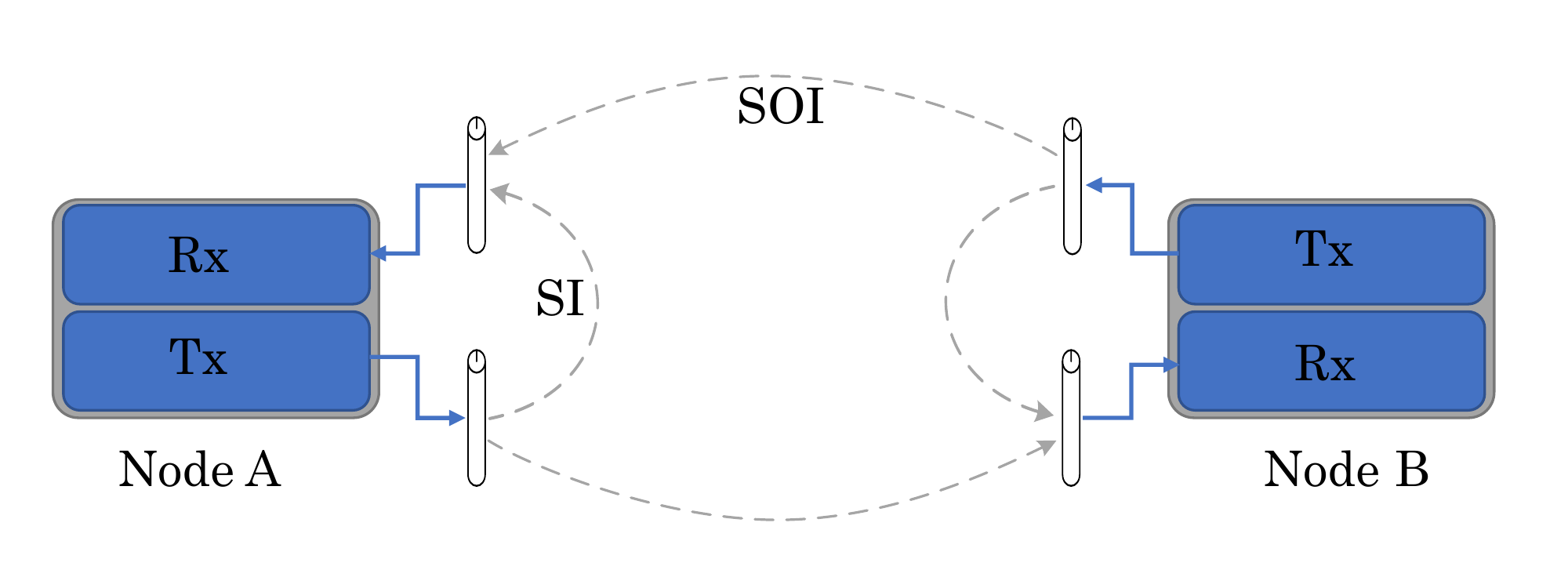}
\vspace{-0.1in}
\caption{Block diagram of FD system.}
\label{FD_diagram}
\vspace{2pt}
\end{figure}

Fig. \ref{FD_diagram} shows the IBFD system which consists of two nodes. Each node is transmitting to and receiving from the other node simultaneously, in the same frequency band. Each node receives a signal of interest (SOI) from the other node and a self-interferer signal (SI) from its own transmitter. Thus, the received signal in the frequency domain can be modeled as
\begin{equation}\label{R_k}
R(k)=\alpha_1 S_{si}(k)+\alpha_2 S_{soi}(k)+ N(k), 
\end{equation}
where $S_{si}(k)$ and $S_{soi}(k)$ are the SI and SOI signals, respectively, associated with subcarrier $k$, $\alpha_1$ and $\alpha_2$ are the complex channel coefficients associated with each, respectively, and $N$ is the complex Additive White Gaussian Noise (AWGN) term in the received signal for the $k-$th subcarrier.

\begin{figure}[!t]
\centering
\subfloat[]{\includegraphics[width=\linewidth]{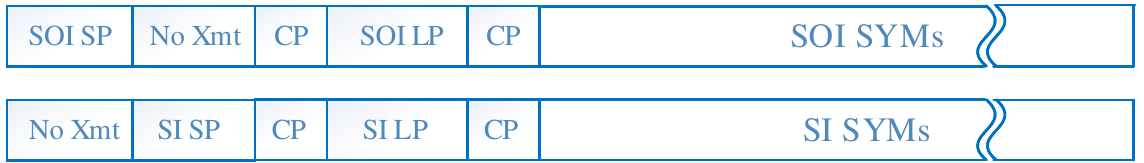}%
\label{OverlapepedLP}}
\vfil
\vspace{-0.05in}
\subfloat[]{\includegraphics[width=\linewidth]{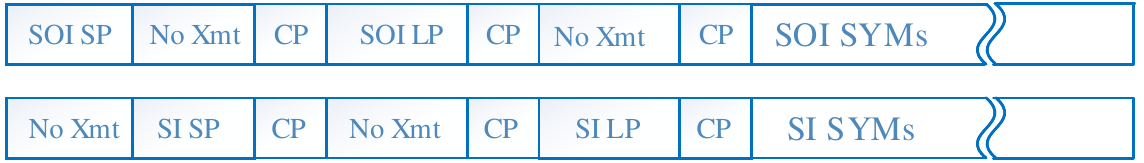}
\label{NonOverlapepedLP}}
\caption{IBFD frame structure (a) with overlapped and (b) with nonoverlapped long premble (LP). SP, CP and NoXmt stand for short preamble, cyclic prefix and no transmission.}
\label{OFDM_frame_fig}
\end{figure}

In order to extract the SOI component from $R(k)$ with high signal-to-interference-plus-noise ratio (SINR), the receiver has to suppress SI close to below the noise floor. Both SOI and SI channels estimation and equalization are needed. A typical approach is to use least squares (LS) estimation, where the receiver relies on the received training sequence with no interferer (i.e nonoverlapped long preamble) and computes the channel estimate as
\begin{equation}\label{LS_equ}
\hat{\alpha}_{1,2}(k)=\sum_{l=1}^{L}\frac{R_l(k)}{T_l(k)}=\alpha_{1,2}+\sum_{l=1}^{L}\frac{N_l(k)}{T_l(k)},
\end{equation}
where $l$ is the training symbol index in time, $T(k)$ is the training or known sequence, and $L$ is the length of the training sequence. This initial estimate is followed by a correction every symbol using nonoverlapped subcarriers (i.e pilots) to track the channel variation.
The channel estimates are used to suppress the SI component, and to equalize SOI channel impact. The estimated SOI can be written as    
\begin{equation}
\hat{S}_{soi}(k)=\frac{R(k)-\hat{\alpha}_{1}(k)S_{si}(k)}{\hat{\alpha}_{2}(k)}.
\end{equation}
\quad In this work, a WiFi-like frame structure is assumed with a nonoverlapped short preamble, shown in Fig. \ref{OFDM_frame_fig}, for IBFD synchronization purposes as discussed in \cite{shaboyan2017frequency}. In order to apply self-interference cancellation, there are two assumptions: a) channel is quasi-static within the frame which is reasonable for low-mobility environments. b) SI and SOI are synchronized to use the knowledge of the interferer signal in the cancellation. Prior work such as \cite{shaboyan2017frequency} present IBFD synchronization techniques that have been experimentally verified, utilizing nonoverlapped short preamble.  

%
\section{ ICA Formulation of IBFD Problem}
In this work, FICA-based BSS is applied in the frequency domain for an Orthogonal Frequency Division Multiplexed (OFDM) system. ICA is applied to the received signal, which is a mixture of SOI and SI in addition to noise. One of the known issues in BSS is the scaling ambiguity problem \cite{comon2010handbook}, where the resulting estimated sources have a multiplicative scaling factor. By taking into account the scaling factor, the problem can be reformulated as follows:
\begin{equation}\label{BSS_MatrixFormulation}
    \begin{bmatrix} X_1(k)\\ X_2(k) \end{bmatrix}=
    \begin{bmatrix} 1/\beta_1&0\\ \alpha_{1}/\beta_1 &\alpha_{2}/\beta_2 \end{bmatrix}
    \begin{bmatrix} \beta_1 S_{si}(k)\\ \beta_2 S_{soi}(k) \end{bmatrix}+
    \begin{bmatrix} 0 \\N(k) \end{bmatrix},
\end{equation}
\noindent where $X_1(k)$ is the direct feed signal (self-interferer) for subcarrier $k$, $X_2(k)$  is the received over the air (mixed) signal, i.e. $R(k)$ of (\ref{R_k}), and $\beta_1$ and $\beta_2$ are the complex scale ambiguity factors associated with the SI and SOI signals, respectively, as perceived by the BSS algorithm. This system can be written as $\boldmath{X(k)=A_aS_a(k)+N(k)=AS(k)+N(k)}$, which is the well known formulation of the noisy BSS problem and where we denote the ambiguated mixing matrix by $\boldmath{A_a}$ and the ambiguated sources vector by $\boldmath{S_a}$. After application of ICA, the estimated sources are recovered, which can be mathematically formulated as $\boldmath{Y=WX=WAS}$ where $\boldmath{W}$ is the demixing matrix. We may write $J=WA$, where $J$ is the estimated coefficient matrix for the sources vector and should ideally be the identity matrix, with $W$  ideally equal to $A^{-1}$, i.e. upon perfect demixing and without ambiguity.
However, due to residual separation errors that depend on many factors including the ICA algorithm used, data size, the number of iterations, and convergence rate, $J$ will have non-diagonal components, where $W$ can be expressed as $\boldmath{W=(A_a^{-1}+ \delta)}$, and $\boldmath{J=(A_a^{-1}+ \delta)A}$, where $\boldmath{\delta}$ is the residual separation error matrix term, which may be written as follows  
\begin{equation}
J=
    \begin{bmatrix} {\beta_1}+\delta_{11}+\delta_{12}\alpha_1&\delta_{12}\alpha_2\\ \delta_{21}+\delta_{22}\alpha_1  & {\beta_2} + \delta_{21} \end{bmatrix}.
\end{equation}
It is evident that if $\{\delta_{ij}\}$, the elements of $\boldmath{\delta}$, are zeros, $J$ is diagonal and if, furthermore, $\beta_1=\beta_2=1$ (i.e no ambiguity scaling), $J=I$.
We are interested in recovering the SOI, which may be written as 
\begin{equation}
    Y_{soi}(k)=J_{21} S_{si}(k) + J_{22} S_{soi}(k)+N_{soi}(k)
    \label{eqYSOI},
\end{equation}
where $\{J_{mn}\}$ are the elements of $J$ and $N_{soi}(k)$ is the AWGN term after applying the demixing matrix $W$. Given known training symbols $S_{soi}(k) = T(k)$, then dividing the LHS by $T(k)$ and taking the sample mean as in (\ref{LS_equ}) yields an estimate for $J_{22}$. This is because $E[(J_{21} S_{si}(k)+ N_{soi}(k))/T(k)]=0$ due to the statistical independence of SOI, SI, and noise terms, thus nulling out the first and third terms of (\ref{eqYSOI}) yielding
\begin{equation}
    J_{22} = E\left[\frac{Y_{soi}(k)}{T(k)}\right].
\end{equation}
With the limited length of the training sequence, the variance of the sample mean estimator for the first term of (\ref{eqYSOI}) may be large due to the much higher power of the SI signal. Hence, using the frame structure presented in \cite{shaboyan2017frequency} and shown in Fig. \ref{OFDM_frame_fig}, non-overlapped Long Preamble (LP) can be used to set SI LP to zero when SOI LP is transmitting.
Having thus obtained the estimate $\hat{J}_{22}$ for the unknown data symbols, we divide $Y_{soi}(k)$ by $\hat{J}_{22}$ to obtain the estimate $\hat{S}_{soi}(k)$ of the source signal for the SOI with good performance as will be shown in the results section.
The permutation ambiguity can also be easily resolved in this case using the same training sequence and the special structure of the demixing matrix afforded by knowledge of the SI signal.


%
\subsection{Modified FICA Algorithm}
BSS works on the statistical independence nature of the sources by either minimizing the mutual information or maximizing the information. Numerous techniques have been proposed for BSS but most assume a real mixing matrix such as FICA, Infomax, and JADA \cite{comon2010handbook}. The problem of SIC in IBFD presents a complex mixing matrix. While complex-ICA methods such as \cite{fu2014complex} exist, they are generally computationally expensive. However, SIC in IBFD can be easily reformulated as a higher dimension BSS real mixing matrix due to the limited dimension of the problem, as we show next.   

FICA is a robust method to extract the independent components and was proposed in \cite{hyvarinen2000independent, hyvarinen1999fast}. There are two methods to separate the signals; the deflection method where the sources are separated one after another, and the symmetric method where the sources are separated simultaneously. In this work, the deflection approach is used due to its ability to estimate a subset of the original components which is the case for IBFD. Since FICA works on real-valued data, we write out the real and imaginary components of the complex mixing problem yielding a 4x4 real matrix formulation. Thus,  (\ref{BSS_MatrixFormulation}) can be rewritten as follows
\begin{equation}
\begin{bmatrix} X_{1r}\\X_{1i}\\X_{2r}\\X_{2i} \end{bmatrix}=
    \begin{bmatrix} 
    1&0&0&0\\0&1&0&0\\
    \alpha_{1r} &-\alpha_{1i} & \alpha_{2r} & -\alpha_{2i}\\
    \alpha_{1i} &\alpha_{1r} & \alpha_{2i} & \alpha_{2r}\\
    \end{bmatrix}
    \begin{bmatrix} S_{si_r}\\S_{si_i}\\S_{soi_r}\\S_{soi_i} \end{bmatrix}+
\begin{bmatrix} 0\\0\\N_r\\N_i \end{bmatrix},\!\!
\end{equation}
where, for convenience, we have dropped the subcarrier indexing and explicit ambiguation, and where the second subscripts $r$ and $i$ denote the real and imaginary parts, respectively.
As in the formulation described earlier, the sample mean estimation is used to obtain the desired coefficient matrix terms, except that two sample mean estimates are needed based on the real and imaginary parts of the training symbols. The deflection algorithm is applied on the last two rows only with identity initial matrix, $\boldmath{W_{i}=I}$,  which satisfies the SI conditions, and using prewhitening and $tanh()$ nonlinearity.  It is worth mentioning that IQ imbalance (IQI) can be lumped inside the mixing matrix which means that the FICA can resolve the IQI while separating the signals (see supplementary material).  
\section{Experimental Results and Discussion}
\subsection{Experiment Setup}
\begin{figure}
    \centering
    \includegraphics[width=0.7\linewidth,height=0.55\linewidth]{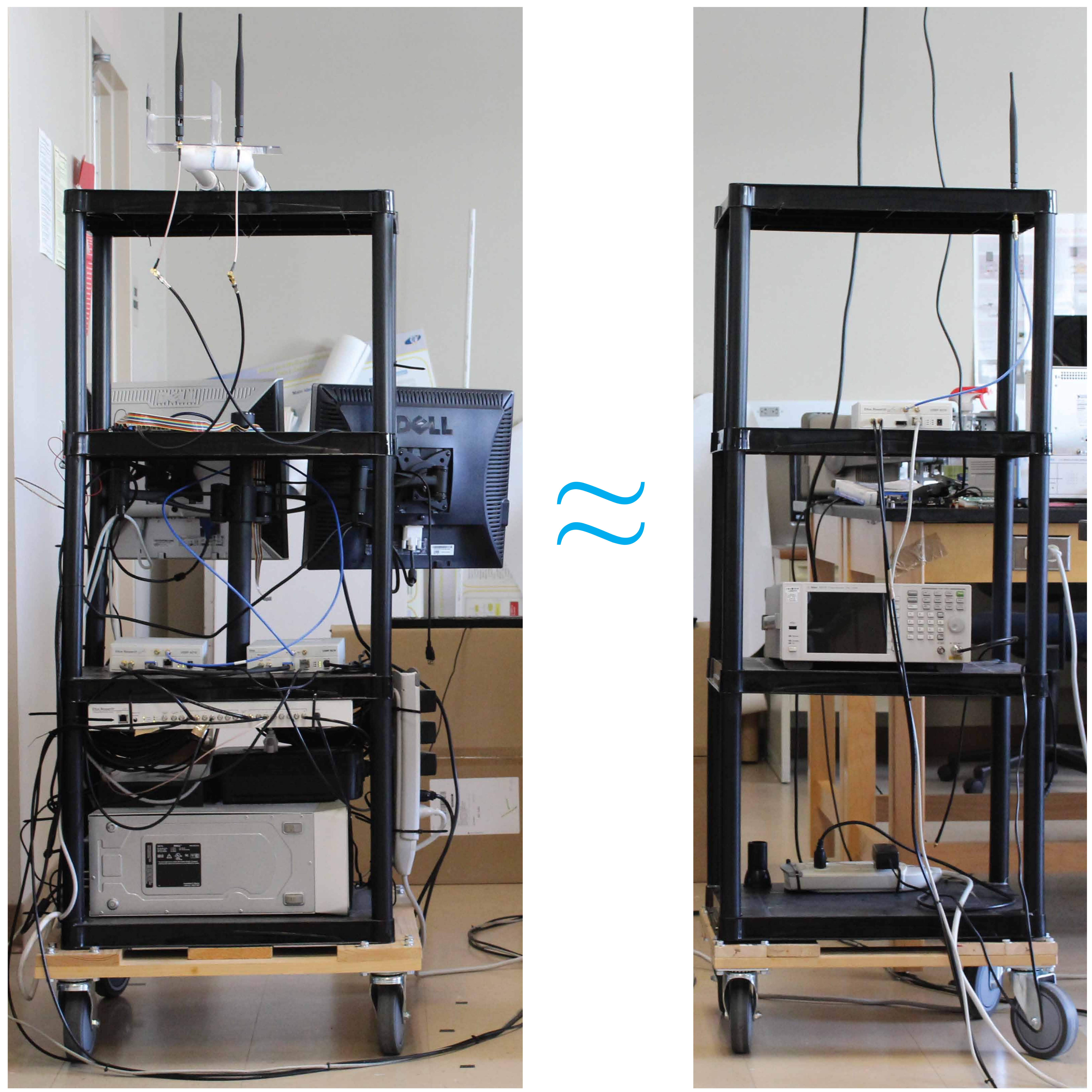}
    \caption{The experimental setup of Node A (left) and Node B (right).}
\label{fig_exp_setup}
\vspace{4pt}
\end{figure}

In order to experimentally validate the proposed SIC method, a complete full-duplex experimental system was built using two Universal Software Radio Peripheral (USRP) SDR platform which contains a Radio Frequency (RF) transceiver and a Field Programmable Gate Array (FPGA). The USRPs are connected to a PC to perform baseband signal processing (mainly DC). The RF transceivers are deployed to perform real-time transmission and reception. The baseband signals are connected to USRPs at a rate of $25M\, Samples/sec$. As shown in Fig. \ref{fig_exp_setup}, Node A is equipped with two dipole omnidirectional antennas; one transmit and one receive antenna with $10\, cm$ separation while Node B uses only one dipole omnidirectional antenna. Both transmit and receive antennas have the same antenna polarization. The distance between the two nodes is set to $10\,m$. The timing of all USRPs and the FPGA that drive the antenna radiation selection are aligned with one reference Pulse Per Second signal. Table \ref{Tab:SimParam} summarizes the parameters used in our framework. The performance metrics, the input signal to interference noise ratio (ISINR), the output signal to interference noise ratio (OSINR) values are calculated at the demodulation step before and after applying DC.

\begin{table}[!t]
\centering
\begin{tabular}{|c|c|c|c|}
\hline
     \textbf{Parameters} & \textbf{Value} & \textbf{Parameters} & \textbf{Value}\\
     \hline
     \# OFDM Subcarriers & 64 & Data Packet Duration & 1.6ms\\
     \# Data Subcarriers & 44/52 & \# symbols/block & 100 \\
     \# Pilot Subcarriers & 8/0 & Carrier Frequency & 2.5 GHz\\
     CP Duration & $3.2\mu s$ & Symbol Duration (CP+FFT) & $16\mu s$\\
     LP Duration & $32\mu s$ & Signal Bandwidth & 5 MHz\\
     SP Duration & $16\mu s$ & Subcarrier Frequency Spacing & 78125Hz\\
     \hline
\end{tabular}
\caption{Experimental setup parameters.}
\label{Tab:SimParam}
\end{table}

\subsection{Comparison with Least-Squares}
We compare FICA-based SIC with LS-based SIC. In LS-based channel estimation, the channel is estimated as discussed in section II in addition to using eight nonoverlapped sub-carriers as pilots; four for each node while the rest are used for data. The estimated channels are used and interpolated through the OFDM symbol to track the changes in each channel with initial estimates using nonoverlapped long preamble. LS-based SIC causes loss of spectral efficiency. On the other hand, FICA-based SIC does not require subcarrier pilots since the whole frame is used to estimate the channel except that the nonoverlapped preamble is needed for ambiguity estimation. Thus, the spectral efficiency of FICA-based SIC is better by $\frac{52}{44} \times$ compared to LS-based SIC where $52$ is the number of used subcarriers in WiFi and $44$ is the number of data subcarriers of LS-based SIC, in our experimental setup.  

It should be noted that the FICA-SIC requires the entire frame for separation whereas the LS-SIC processes the frame symbols sequentially,  i.e. the latency of FICA-SIC is O(N), relative to O(1) for LS-SIC which serves other functions such as decoding.  

\subsection{Experiment Results}
\begin{figure}[!t]
\centering
\includegraphics[width=\linewidth]{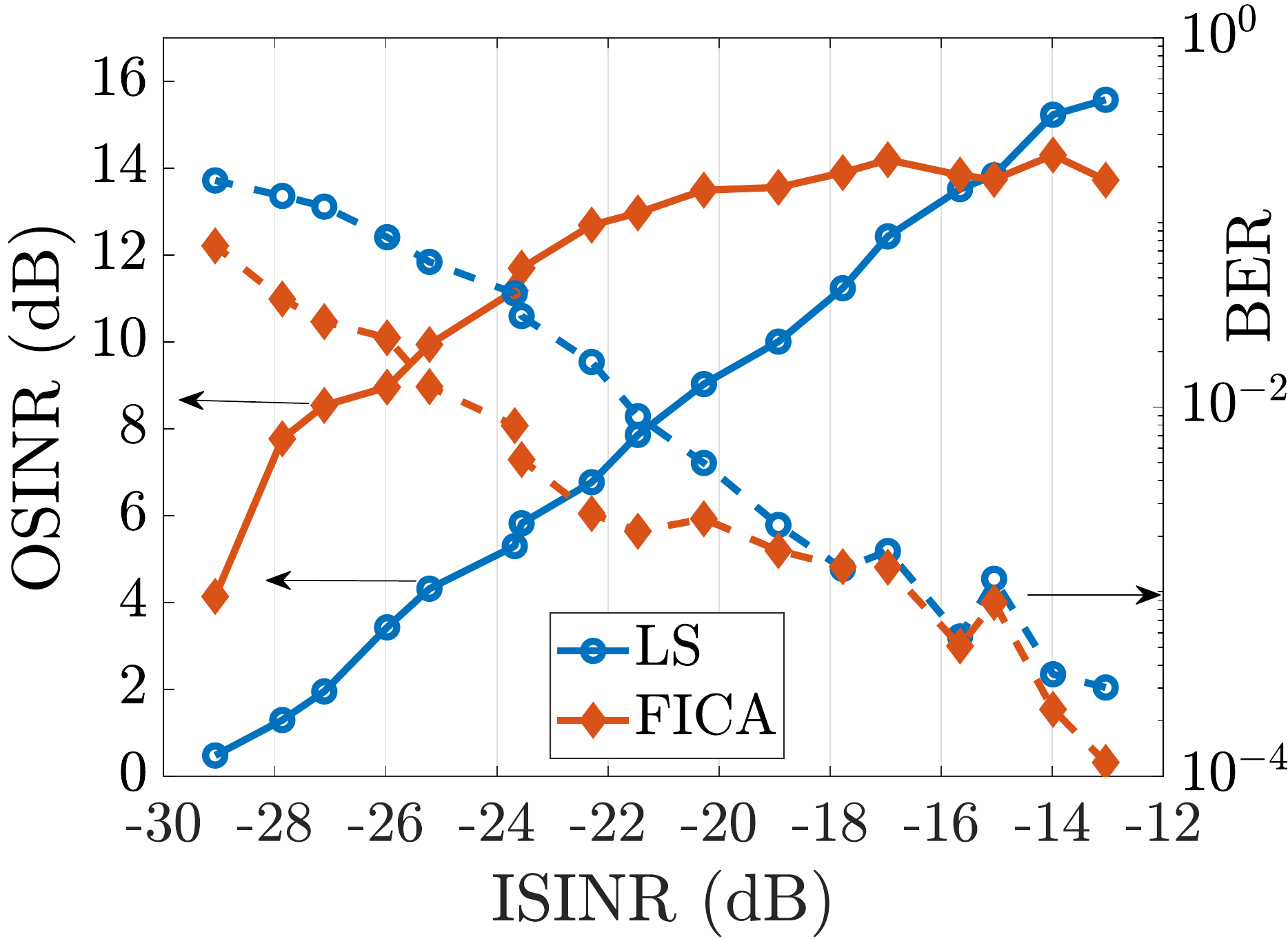}%
\caption{Experimental comparison of the output SINR (solid lines) and bit error rate (dashed lines) for different input SINR.}
\label{OSINR_ISINR}
\end{figure}

\begin{figure}[!t]
\centering
\includegraphics[width=\linewidth]{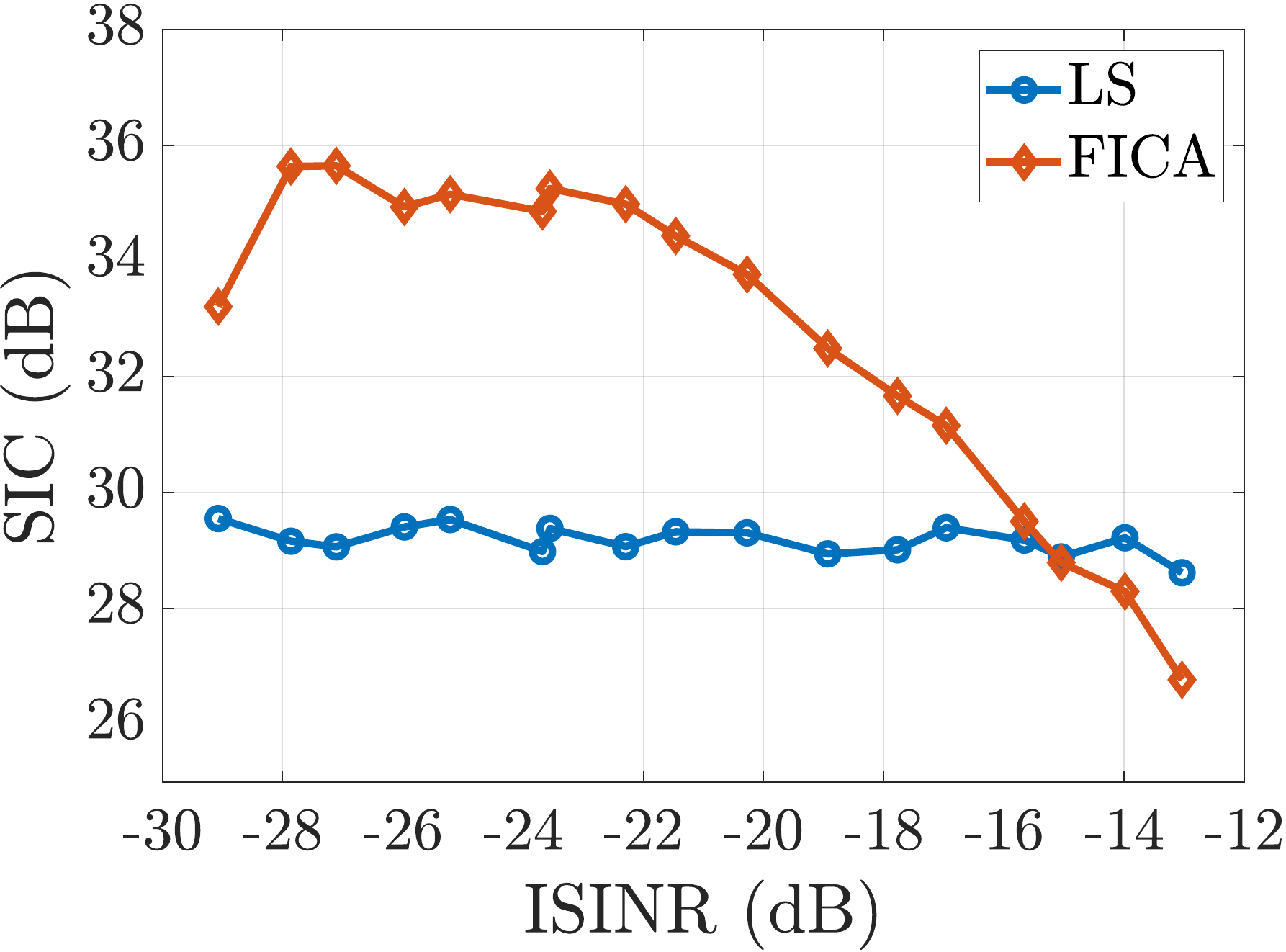}%
\caption{Experimental comparison of the SIC for different input SINR between LS- and FICA-based SIC.}
\label{SIC_ISINR}
\end{figure}

In order to sweep over ISINR, the SOI Tx power is swept from 0 dBm up to 15 dBm with a 0.5 dBm step while fixing the SI Tx power to 0 dBm. The measured SI power before DC is fixed to -37 dBm. The performance after applying both FICA-based and LS-based cancellation are shown in Fig. \ref{OSINR_ISINR}. Also, the overall SIC is depicted in Fig. \ref{SIC_ISINR}. Clearly, FICA-based cancellation outperforms the LS-based cancellation by around 6 dB due to the intrinsic de-noising feature in  ICA\cite{comon2010handbook}. However, the performance starts to saturate due to the nonlinearity of both transmitter and receiver. FICA-based cancellation is more sensitive to the nonlinearity.          


\begin{figure}[!t]
\centering
\includegraphics[width=\linewidth]{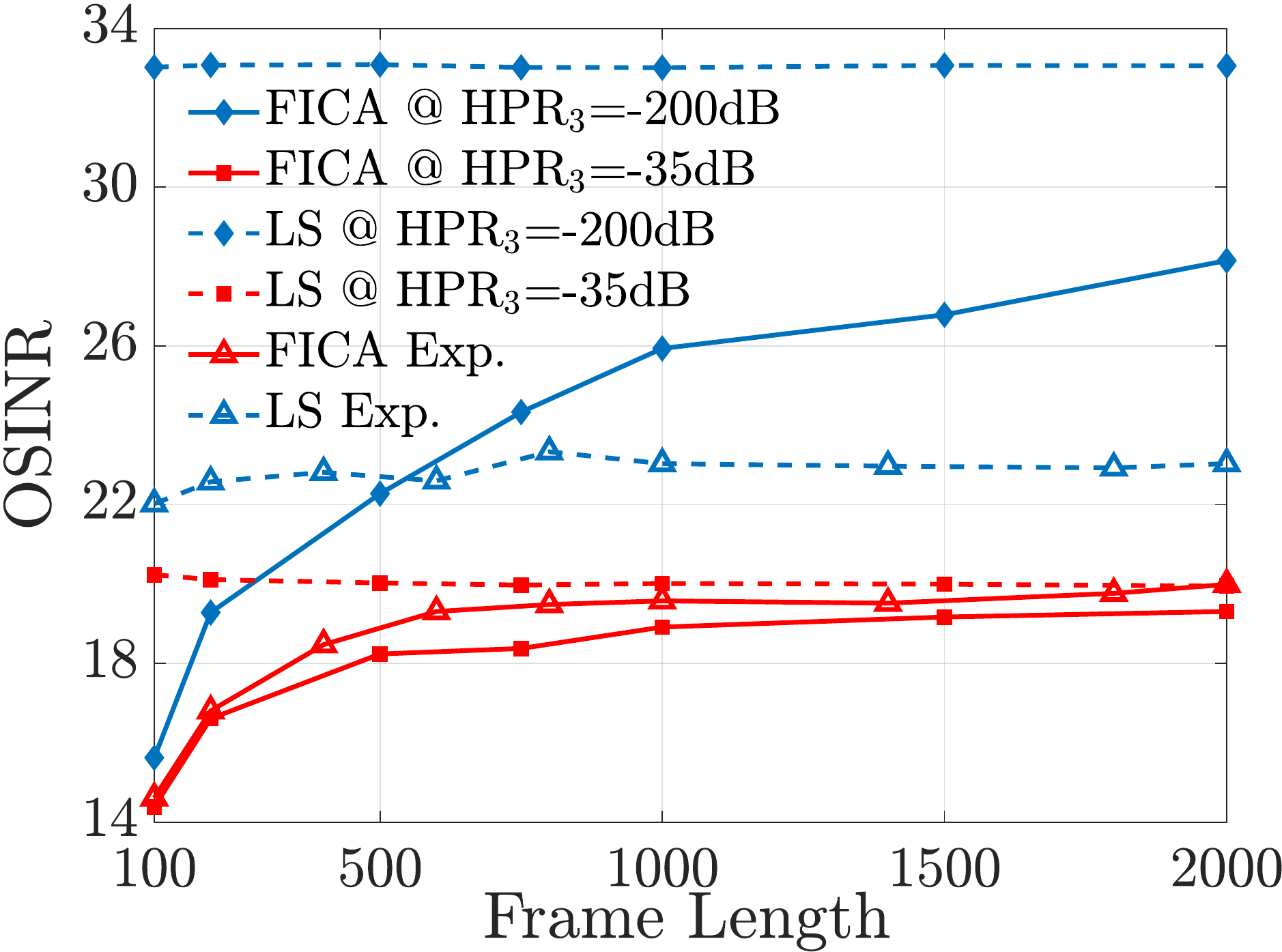}
\vspace{-0.1in}
\caption{Experimental and simulation of the OSINR for different frame lengths with ISINR=-10 dB.}
\label{OSINR_N}
\end{figure}

In order to show the effect of the nonlinearity,  experiments with different third harmonic power ratio (HPR${_3}$) were conducted in the simulation platform and verified experimentally for different frame lengths. Fig. \ref{OSINR_N} shows LS-based and FICA-based SIC for HPR${}_3=200\,$ dB representing no nonlinearity and HPR${}_3=35\,$ dB representing the practical values existing in our experimental framework and compared to the experimental result. Clearly, the nonlinearity causes performance saturation. This occurs due to the assumption of the linear mixture which is true only for low ISINR. Thus, nonlinearity estimation techniques can be used to improve the performance or nonlinear ICA techniques can be applied \cite{comon2010handbook}, which may be the subject of future work.

\section{Conclusion and Future Work}
In this letter, ICA is deployed to solve the self-interference problem in IBFD systems in the frequency domain of an OFDM system. The self-interference problem formulation as an ICA problem is introduced. Experimental results show that the proposed technique is very effective in solving the SI problem thanks to the intrinsic de-noising feature in ICA. Besides, the FICA-based SIC achieves spectral efficiency improvement by an extra 18\% compared to LS-based SIC. Comparisons to other SIC schemes would be of interest as well. Another natural extension to this work would be application of BSS to SIC in IBFD-MIMO since multiple antennas would add new received signals with independent channels to the linear mixture. The specific BSS problem formulation in this case would depend on the diversity-multiplexing tradeoff used. Beyond the obvious complexity issue, there are a few challenges as well as opportunities associated with multiple antennas \cite{Kim2015, Nguyen2018}. These include the design of MAC protocols to minimize network interference (and, in our proposed scheme, to coordinate the synchronization of preambles), the need for stronger SIC algorithms, precoder design, and the need for theoretical performance bounds on the capacity increase for multiple shared antennas. 



\ifCLASSOPTIONcaptionsoff
  \newpage
\fi

\end{document}